\documentclass[twoside]{article}
\usepackage{fleqn,espcrc2}
\usepackage{graphicx}
\usepackage{here}

\newcommand{\AmS}{{\protect\the\textfont2
    A\kern-.1667em\lower.5ex\hbox{M}\kern-.125emS}}
\def\beq{\begin{equation}}
\def\eeq{\end{equation}}
\def\bea{\begin{eqnarray}}
\def\eea{\end{eqnarray}}
\def\bq{\begin{quote}}
\def\eq{\end{quote}}

\def\nnb{\nonumber}
\def\ga{\left(}
\def\dr{\right)}

\def\rar{\rightarrow}

\def\nnb{\nonumber}
\def\la{\langle}
\def\ra{\rangle}
\def\nin{\noindent}
\def\ba{\begin{array}}
\def\ea{\end{array}}

\def\b{$\bullet$}
\def\als{\alpha_s}

\def\gg2{ \la\alpha_s G^2 \ra}
\def\gg3{g^3f_{abc}\la G^aG^bG^c \ra}
\def\ggg4{\la\als^2G^4\ra}

\title{\bf{\boldmath
{\huge Scalar mesons and the muon anomaly} }}
\author{
Stephan Narison\address{ Laboratoire
de Physique Math\'ematique et Th\'eorique, Universit\'e
de Montpellier II, Case 070, Place Eug\`ene
Bataillon, 34095 - Montpellier Cedex 05,
France.\\ E-mail:
qcd@lpm.univ-montp2.fr} }
\begin{document}
\textwidth 17cm
\textheight 24.5cm
\topmargin -2.5cm
\oddsidemargin -.6cm
\evensidemargin -0.5cm
\pagestyle{empty}
\pagestyle{plain}
\begin{abstract}
\noindent
We evaluate systematically some new contributions of the QCD scalar
mesons, including radiative decay-productions,  not considered with a
better attention
until now in the evaluation of the hadronic contributions to the muon
anomaly. The sum of the scalar contributions to be added to the
existing Standard Model predictions
$a_\mu^{SM}$ are estimated in units $10^{-10}$ to be
$ a^{S}_\mu= 1.0(0.6) ~\big{[}{\rm TH~based}\big{]}$ and $13(11) ~\big{[}{\rm
PDG~based}\big{]}$, where the errors are  dominated by the ones
from the experimental widths of these scalar mesons. PDG based results
suggest that the value of $a_\mu^{SM}$
and its errors might have been underestimated in previous works. The
inclusion of these
new effects leads to a perfect  agreement ($\leq 1.1\sigma$) of the measured
value
$a^{exp}_\mu$ and
$a_\mu^{SM}$ from
$\tau$-decay  and implies a $(1.5\sim 3.3)~\sigma$ discrepancy between
$a^{exp}_\mu$ and $a_\mu^{SM}$ from $e^+e^-\rar$ hadrons data.
More refined unbiased estimates of $a_\mu^{SM}$ require improved
measurements of the scalar meson masses and widths. The impact of our
results to
$a_\mu^{SM}$ is summarized in the conclusions.
\end{abstract}
\maketitle
\section{INTRODUCTION}
\nin
Improved accurate experimental measurement \cite{BNL} and recent
theoretical estimates of the muon
anomaly \cite{DAVIER,KAORU,SN,YND} are now available. The
theoretical accuracy in \cite{DAVIER,KAORU}~\footnote{Though agreeing
in the total sum, the results from
these two estimates  differ in each energy region.} is mainly
attributed to the use of the new
CMD-2
\cite{CMD2} around the
$\rho$ mass and on BES data \cite{BEPC} around the $J/\Psi$ region.
The impact of the former data on the determination of
the muon anomaly is intuitively more important due to the low-energy
dominance of the anomaly kernel function
\cite{RAF,LAUT}. However, what is more intriguing within this
increasing precision is the discrepancy
between the results from
$e^+e^-$ and $\tau$-decay data \cite{DAVIER,KAORU}, which was not the
case in the previous determinations
using preliminary data
\cite{SN,YND}. In this short note, we study some other sources of
contributions, not considered with
a better attention until now, from the scalar mesons. These  scalar
($\bar qq$, gluonia,...) are
conceptually fundamental consequences of QCD. Though their existence
is not precisely confirmed, there are increasing
evidences of their findings in different $e^+e^-$ and hadronic
experiments \cite{KYOTO}.
\section{ISOSCALAR SCALAR MESONS}
\nin
The interest in these $I=0$ scalar mesons are that they cannot be
obtained from usual ChPT approaches.
They are related to the QCD scale anomaly:
\beq\label{eq:anomaly}
\theta_\mu^\mu=\frac{1}{4}\beta(\alpha_s) G^2+\sum_i [ 1+\gamma_m(\alpha_s)]
m_i\bar\psi_i\psi_i~,
\eeq
where $G^a_{\mu\nu}$ is the gluon field strengths, $\psi_i$ is the
quark field; $\beta(\alpha_s)$ and
$\gamma_m(\alpha_s)$ are respectively the QCD $\beta$-function and
quark mass-anomalous dimension.
In this case, arguments based on $SU(2)$ symmetry or its violation
used to estimate some
processes like e.g. radiative processes cannot be applied \cite{U1}.
Using QCD spectral sum rules (QSSR)
\cite{SVZ,SNB} and low-energy theorems (LET) for estimating the mass
and its width, it has been shown in
\cite{VENEZIA,SNG} that the wide
$\sigma(0.60)$ meson is the best candidate meson (gluonium)
associated to the previous interpolating current (see also
\cite{OCHS}). Its mass
is due to the gluon component and its large width into $\pi\pi$ is
due to a large violation of the OZI rule \footnote{This feature also
implies that lattice
calculations of the gluonium mass in pure Yang-Mills or quenched
approximation are bad approximations and might miss
this "unusual" glueball.} (analogue
of the
$\eta'$-meson of the $U(1)_A$ sector \cite{WITTEN}).  The oberved
$\sigma(0.60)$ and $f_0(0.98)$
can be explained by a ``maximal quarkonium-gluonium mixing scheme"
\cite{VENEZIA,SNG}. In the following, we study the $\sigma$ and other
scalar mesons
contributions to the muon anomaly.
\section{THE $\bf e^+e^-\rar ~S\gamma$ PROCESSES}
\nin
Using analytic properties of the photon propagator, the general
contribution to the muon anomaly from the process:
\beq
e^+e^-\rar \gamma^*\rar hadrons
\eeq
can be written in a closed form as \cite{RAF}:
\bea
a_l^{had}(l.o)=\frac{1}{4\pi^3}\int_{4m^2_\pi}^\infty dt~K_l(t)~\sigma_H(t)~,
\eea
where:\\
\b~$K_l(t\geq 0)$ is the QED kernel function \cite{LAUT}:
\bea\label{kernel}
K_l(t)&=&\int_0^1 dx\frac{x^2(1-x)}{x^2+\ga{t}/{m_l^2}\dr(1-x)}~,
\eea
with the analytic form:
\bea
K_l(t\geq 4m_l^2)&=&
z_l^2\ga 1-\frac{z_l^2}{2}\dr+
\ga
1+z_l\dr^2\times\nnb\\
&&\ga 1+
\frac{1}{z_l^2}\dr\Big{[}\log(1+z_l)-z_l\nnb\\
&&+\frac{z_l^2}{2}\Big{]}
+\ga\frac{1+z_l}{1-z_l}\dr z_l^2\log{z_l},
\eea
with:
\beq
y_l=\frac{t}{4m^2_l},~~z_l=\frac{(1-v_l)}{(1+v_l)}~~{\rm and
}~~v_l=\sqrt{1-\frac{4m_l^2}{t}}.
\eeq
   $K_l(t)$ is a monotonically decreasing function of $t$. For large $t$,
it behaves as:
\beq
K_l(t> m^2_l)\simeq \frac{m^2_l}{3t}~,
\eeq
which will be an useful approximation for the analysis in the large $t$ regime.
Such properties then emphasize the importance of the
low-energy contribution to $a_l^{had}(l.o)~(l\equiv e,~\mu)$, where
the QCD analytic calculations cannot
be applied.\\
\b~$\sigma_H(t)\equiv\sigma(e^+e^-\rar{\rm
hadrons})$ is the $e^+e^-\rar $ hadrons total cross-section which can be
related to the hadronic two-point spectral function Im$\Pi(t)_{em}$ through the
optical theorem:
\beq
R_{e^+e^-}\equiv\frac{\sigma(e^+e^-\rar{\rm
hadrons})}{\sigma(e^+e^-\rar\mu^+\mu^-)}=12\pi{\rm Im}\Pi(t)_{em}~,
\eeq
where:
\beq
\sigma(e^+e^-\rar\mu^+\mu^-)=\frac{4\pi\alpha^2}{3t}.
\eeq
Here,
\bea\label{twopoint}
\Pi^{\mu\nu}_{em} &\equiv& i \int d^4x ~e^{iqx} \
\la 0\vert {\cal T}
{J^\mu_{em}(x)}
\ga {J^\nu_{em}(x)}\dr^\dagger \vert 0 \ra \nnb\\
&=&-\ga g^{\mu\nu}q^2-q^\mu q^\nu\dr\Pi_{em}(q^2)
\eea
is the correlator built from the local electromagnetic current:
\beq
J^\mu_{em}(x)=\frac{2}{3}\bar u\gamma^\mu u-
\frac{1}{3}\bar d\gamma^\mu d-\frac{1}{3}\bar s\gamma^\mu s+...
\eeq
Using Vector Meson Dominance Model (VDM) in a Narrow Width
Approximation (NWA), one obtains
\cite{SNTHESIS}:
\beq
a_\mu^{VDM}(l.o)\simeq {3\over\pi}K\ga{M^2_V\over
m^2_l}\dr{\Gamma_{V\rar e^+e^-}\over M_V}{\Gamma_{V\rar \gamma X}\over
\Gamma_{V\rar all}}~.
\eeq
The $V\rar \gamma X$ ($X\equiv \sigma, f_0$) coupling has been
estimated from the $X\gamma\gamma$ one given in
\cite{VENEZIA,SNG} and in the PDG data \cite{PDG} and using the VDM relation:
\beq
g_{X\gamma V}\simeq {\sqrt{2}\gamma_V\over e} g_{X\gamma\gamma}~,
\eeq
where $e$ is the electric charge and we use the normalization
$\gamma_\rho\simeq 2.51\pm 0.02$. For the scalar
mesons, we use:
\bea
M_\sigma&\simeq& (0.6\sim 0.8)~{\rm GeV}~,\nnb\\
\Gamma_{\sigma\rar\gamma\gamma}&\simeq& (0.2\sim 0.3)~{\rm
keV~ QSSR}~\cite{VENEZIA,SNG}\nnb\\
&\simeq&(3.8\pm 1.5)~{\rm
keV~~PDG}~\cite{PDG,MENES}~,\nnb\\
\Gamma_{f_0(1.4)\rar\gamma\gamma}&\simeq&
(0.7\sim 5.)~{\rm keV}~,
\eea
  The ones of the
other
$\omega$ and
$\phi$ radiative widths come from PDG
\cite{PDG}. We also use in MeV units
\cite{PDG}:
\bea
\Gamma_{\omega(1.42)\rar e^+e^-}&\simeq& 0.08\times
10^{-3},~\Gamma_{tot}\simeq 174(59)\nnb\\
\Gamma_{\rho(1.45)\rar e^+e^-}&\simeq &0.44\times
10^{-3},~\Gamma_{tot}\simeq 310(60)\nnb\\
\Gamma_{\phi(1.68)\rar e^+e^-}&\simeq& 0.48\times
10^{-3},~\Gamma_{tot}\simeq 150(50)~,\nnb\\
\eea
We deduce the
results in Table \ref{tab:table1},
where the first set of values corresponds to
$\Gamma_{\sigma\rar\gamma\gamma}$ from QSSR
and the second set of values to the $\sigma$ width from PDG.
\begin{table}[hbt]

\setlength{\tabcolsep}{0.7pc}
\caption{$e^+e^-\rar~ $scalar+$\gamma$ contributions to $a_\mu^{had}$ $^{*)}$.
}
\label{tab:table1}
\begin{tabular}{llll}
\hline
&\\
{\bf Processes}&{$\bf a_\mu^{had}\times 10^{10}$}&\\ &&\\
\hline
&&\\
& QSSR&PDG\\
&&\\
$\rho\rar\sigma\gamma$&$\sim 0.$&$\sim 0.05$ \\
$\omega\rar\sigma\gamma$&$\sim 0.$&$\sim 0.07$& \\
$\phi\rar\sigma\gamma$&$0.03\sim 0.15$&$1.\sim 10.$& \\
$\phi\rar f_0(.98)\gamma$&PDG&$0.01\pm 0.00$& \\
$\phi\rar a_0(.98)\gamma$&PDG&$0.03\pm 0.01$& \\
\\
$\rho(1.45)\rar\sigma\gamma$&$0.01\sim 0.02$&$0.06\sim 0.42$& \\
$\omega(1.42)\rar\sigma\gamma$&$0.10\sim 0.17$ &$0.1\sim 0.7$\\
$\phi(1.68)\rar\sigma\gamma$&$0.14\sim 0.20$&$0.17\sim 1.17$&\\
$\phi(1.68)\rar f_0(1.4)\gamma$&$-$&$-$\\

\\
Total &$0.45\pm 0.13$&$7.3\pm 5.6$\\
&&\\
\hline
\end{tabular}
{\footnotesize
\begin{quote}
$^{*)}$\,For completeness, we also include the contribution of the
isovector $a_0(980)$.
\noindent
\end{quote}}
\end{table}
\nin
It is clear from Table \ref{tab:table1} that the results are very
sensitive to the value of the $\sigma$ mass and $\gamma\gamma$ width.
Though, in the ``maximal quark-gluonium mixing scheme" of the
$\sigma$ meson, one favours a small $\gamma\gamma$ width
\cite{VENEZIA,SNG}, the PDG
data \cite{PDG,MENES} and some other QCD models still allow higher
values. For a conservative and unbiased estimate, we translate the
total sum in Table
\ref{tab:table1} into:
\bea
a_\mu^{S\gamma}\times 10^{10}&=&0.45(0.13)~~~ {\rm QSSR}\nnb\\
&=& 7.30(5.60)~~~{\rm PDG}~.
\eea
In order to include this result into the one in \cite{DAVIER}, and
for avoiding an eventual double counting, we ``de-rescale" their
result on the top of the
$\omega$ by 0.888 and on the $\phi$ by the factor 0.984 used there
for taking into account the missing modes. In this way, we obtain the
corresponding ``pure" hadronic decay:
\beq
a_\mu(\omega+\phi\rar 3\pi+ K\bar K)= 67.04(1.50)\times 10^{-10}~,
\eeq
Using the $\omega$ and $\phi$ into $\eta\gamma$ and $\pi^0\gamma$
radiative decay widths from \cite{PDG}, we deduce:
\beq
a_\mu(\omega+\phi\rar \eta\gamma+\pi^0\gamma)=4.36(0.16)\times 10^{-10}~.
\eeq
Taking into account these different effects, we conclude from this
analysis that the inclusion of these new radiative decays
increases the results in \cite{DAVIER} by:
\bea\label{eq:axg}
\Delta a_\mu^{X\gamma}\times 10^{10}&=&0.10(0.13)~~~ {\rm QSSR}\nnb\\
&=& 6.95(5.60)~~~{\rm PDG}~.
\eea
PDG data for the widths of the $\sigma$ meson may indicate that the 
scaling factors used
in
\cite{DAVIER} are  not sufficient  for taking into account the 
different missing modes
of the
$\omega$ and $\phi$ mesons.
   Adding our result in Eq. (\ref{eq:axg}), to
the recent estimate of $a_\mu^{had}(l.o)$ in
\cite{DAVIER,KAORU,SN,YND}, the lowest order hadronic contributions
to $a_\mu$ becomes~\footnote{For the $\tau$-decay data,
we use the average of the results (in units of 10$^{-10}$)
$709.0(5.9)$ \cite{DAVIER} and
$703.6(7.6)$ \cite{SN}. For the
$e^+e^-$ data, we use the average of the most recent estimates
684.0(6.5) \cite{DAVIER} and 683.1(6.2) \cite{KAORU}, which are
however lower by about $1.4\sigma$ than the previous estimates in
\cite{SN,YND} based on preliminary data.} in units of $ 10^{-10}$:
\bea\label{eq:had}
a_\mu^{had}(l.o)&=&
706.4(6.8)(0.1)~\big{[}\tau-\rm QSSR\big{]}~,\nnb\\
&=&713.3(6.8)(5.6)~\big{[}\tau-\rm PDG\big{]}~,\nnb\\ &=&
683.7(6.4)(0.1)~\big{[}e^+e^--\rm QSSR\big{]}~,\nnb\\
&=& 690.6(6.4)(5.6)~\big{[}e^+e^--\rm PDG\big{]}~,
\eea
where the first error is the ones from \cite{DAVIER,KAORU,SN}, and
the second one comes from the present analysis.
Our
results, especially the one based on PDG data, suggest that the estimate of
$a_\mu^{had}(l.o)$ and its  errors might have been underestimated in 
the previous
determinations.
\section{NEW CONTRIBUTIONS OF THE  SCALAR MESON TO $a_\mu$}
\begin{figure}[hbt]
\begin{center}
\includegraphics[width=3cm]{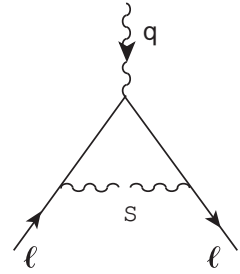}
\caption{Scalar meson contribution to $a_l$.}
\label{fig:scalar}
\end{center}
\end{figure}

\nin
Here, we study the effect of scalar mesons via a Higgs-like triangle
diagram shown in Fig.~\ref{fig:scalar}. This new hadronic
contribution to $a_\mu^{had}$
cannot be estimated using the usual electromagnetic spectral function
associated to one virtual photon.
\nin
The evaluation
of this diagram gives the contribution (see e.g. \cite{CALMET}):
\beq
a_\mu^{Sll}={|g_F|^2\over 16\pi^2}2\int_0^1 dx{x^2(2-x)\over x^2+\ga
M^2_S/ m^2_\mu\dr(1-x)}~,
\eeq
where $M_S$ is the scalar mass and $g_F$ is the ``effective"
$Sl^+l^-$ coupling.
However, the coupling of these scalar mesons to
$e^+e^-$ is not known, and in the case of the isoscalar, the na\"\i
ve argument based on chiral symmetry (Higgs-type
coupling) may not be applied due to the presence of gluon fields in the
$U(1)_V$ dilaton current given in Eq.~(\ref{eq:anomaly}). {Note that
these scalar mesons might also be
produced in the $s$-channel in $e^+e^-$ experiments, most probably
via two-photon exchange, which should
differ from the usual spectral function of a one photon exchange used
to estimate
$a_\mu^{had}(l.o)$, and therefore prevent from some eventual double
counting of the $e^+e^-\rar$ hadrons data. However, we are not also
aware of existing
analyses of the angular  distribution which can differentiate a
scalar from a vector meson in the low-energy
$e^+e^-$ region concerned here
\cite{RENARD}, while present LEP limits on electroweak scalar
particles (sleptons, squarks) given in
\cite{SUZY} may not be applied here.}.
\begin{table}[hbt]
\setlength{\tabcolsep}{0.9pc}
\caption{Contribution of Fig. \ref{fig:scalar} to $a_\mu$ using the
upper bound on the leptonic widths compiled by PDG~\cite{PDG}.}
\label{tab:table2}
\begin{tabular}{lll}
\hline
&\\
{\bf }&{$\bf a_\mu\times 10^{10}$}&$\bf \Gamma_{S\rar e^+e^-}$ \bf [eV]\\
&&\\
\hline
&&\\
$\sigma(.60)$&$0.\sim 8.9$&$\leq 20~^{*)}$ \\
$f_0(.98)$&$0.\sim 1.1$&$\leq 8.4$ \\
$a_0(.98)$&$0.\sim 0.2$&$\leq 1.5$\\
$f_0(1.37)$&$0\sim 1.1$&$\leq 20$\\
&&\\
Total&$0.\sim 11.4$&\\

   &&\\
\hline
\end{tabular}
{\footnotesize
\begin{quote}
$^{*)}$\,Due to the non-available experimental leptonic width of the $\sigma$
and the unclear separation between the $\sigma$ and the $f_0(1.37)$, one can
safely assume that the bounds for the
$\sigma$ and for the $f_0(1.37)$ are about the same.
\noindent
\end{quote}}
\end{table}
\nin
If one estimates this coupling from the
experimental bound on
the $S\rar e^+e^-$ width
\cite{PDG} \footnote{Note that the difference
$a_e^{exp}-a_e^{SM}=(34.9\pm 28.4)\times 10^{-12}$ from
the electron anomaly \cite{MARCIANO} leads to a weaker upper bound.},
and use the positivity of the contribution, one gets the
result in Table
\ref{tab:table2}, where an universal coupling  to $l^+l^-$ ($l\equiv
e,\mu$) has been assumed~\footnote{The bound is weaker for
a Higgs-type coupling.}. In order to see the strength of this
experimental bound, one may assume that the coupling of the scalar to
$l^+l^-$ is dominated by the
one through two photons. Therefore, an {\it alternative rough
estimate} of this effect can be obtained by relating it to
the light-by-light (LL) scattering diagram contribution \cite{LL},
where we obtain  \footnote{We use the total LL contribution in order
to take into
account other mesons contributions}:
\beq\label{eq:aee}
a_\mu^{Sll}({LL})\approx e^2a_\mu^{LL}\approx  (0.1\sim 1.0)\times 10^{-10}~,
\eeq
where we have considered the uncertainties to be about one order of 
magnitude in
order to have a conservative estimate. This result indicates that the present
experimental upper bound in Table~\ref{tab:table2} might  be very weak,
and needs to be improved. We translate the results given in Eqs. (\ref{eq:aee})
and Table~\ref{tab:table2} into (in units of $10^-10$):
\bea\label{eq:ae+e-}
a_\mu^{Sll}&=& 0.55(0.45)~{\rm TH}\nnb\\
&=&5.7(5.7)~{\rm PDG}.
\eea
Adding the results in Eqs. (\ref{eq:axg}) and (\ref{eq:ae+e-}), one
finally deduces the sum of the additionnal contributions due to
scalar mesons in units of $10^{-10}$:
\bea\label{eq:scalar}
a^{S}_\mu&=& 1.0(0.6)~{\rm TH}\nnb\\
&=&13(11)~{\rm PDG}~,
\eea
to be added to the SM predictions $a_\mu^{SM}$. One should notice that the
present theoretical estimate based on QCD spectral sum rules (QSSR) differs by
about an order magnitude to the one based on the PDG data for the $\sigma
\gamma\gamma$ and scalar meson leptonic widths. The QSSR estimate of the
$\sigma\gamma\gamma$ width is based on the picture where the observed $\sigma$
meson emerges from a maximal mixing between a gluonium and $\bar qq$ states,
which then implies a much smaller $\gamma\gamma$ width. On the other, the
theoretical estimate of the leptonic width is based on an intermediate
$\gamma\gamma$ coupling of the scalar meson to the lepton pair explaining again
the relative suppression of a such width.
A progress in improving the accuracy of the scalar meson contributions needs
solely improved measurements of the scalar mesons $\gamma\gamma$ and leptonic
widths. In addition, such a program is also necessary to clarify the exact
nature of the scalar mesons which, at present, is, experimentally, 
poorly known.
\section{CONCLUSIONS}
\nin
The inclusion of the scalar meson contribution in Eq.~(\ref{eq:axg})
modifies the recent estimate of
$a_\mu^{had}(l.o)$ from \cite{DAVIER,KAORU,SN} into the one in
Eq.~(\ref{eq:had}), while the direct exchange of the scalar
meson shown in Fig.~\ref{fig:scalar} gives the new contribution in
Eq.~(\ref{eq:ae+e-}). The total sum of the scalar contributions
is given in Eq. (\ref{eq:scalar}).
   Adding our result in Eqs.~(\ref{eq:had}) and (\ref{eq:ae+e-}), we
deduce in units of $10^{-10}$:
\bea
a_\mu^{had}&\equiv& a_\mu^{had}(l.o)+a_\mu^{Sll}\nnb\\
&=& 707.4(6.8)(0.6)~\big{[}\tau-{\rm TH}\big{]}~,\nnb\\
&=& 719.4(6.8)(11)~\big{[}\tau-{\rm PDG}\big{]}~,\nnb\\
&=& 684.6(6.4)(0.6)~\big{[}e^+e^--{\rm TH}\big{]}~,\nnb\\
&=& 696.6(6.4)(11)~\big{[}e^+e^--{\rm PDG}\big{]}~,
\eea
where the first error is the ones from \cite{DAVIER,KAORU,SN}, and
the second one comes from the present analysis. Our
results (especially the one using PDG data) suggest that the value 
and errors of
the hadronic  contributions to $a_\mu^{SM}$ might have been underestimated in
the  previous
determinations. From the above results, we deduce our final estimate
of the muon anomaly in units of $10^{-10}$:
\bea
a_\mu^{ SM}&=&11~659~191.6(6.8)~~\big{[}\tau-{\rm
TH}\big{]}~,\nnb\\
&=&11~659~203.6(12.9)~\big{[}\tau-{\rm
PDG}\big{]}~,\nnb\\
&=&11~659~169.2(6.4)~~\big{[}e^+e^--{\rm TH}\big{]}~,\nnb\\
&=&11~659~181.2(12.7)~\big{[}e^+e^--{\rm PDG}\big{]}~,
\eea
compared with the recent data \cite{BNL}:
\bea
a_\mu^{exp}&=&11~659~204.0(8.6)\times 10^{-10}~.
\eea
Then, we deduce  in units of $10^{-10}$:
\bea
a_\mu^{exp}-a_\mu^{
SM}&=&12.4(11.0)~\big{[}\tau-{\rm TH}\big{]}~,\nnb\\
&=&0.4(15.5)~\big{[}\tau-{\rm PDG}\big{]}~,\nnb\\
&=&34.8(10.7)~\big{[}e^+e^--{\rm TH}\big{]}~,\nnb\\
&=&22.8(15.3)~\big{[}e^+e^--{\rm PDG}\big{]}~.
\eea
One can see that the inclusion of these new effects due to scalar 
mesons improves
the agreement between $a_\mu^{exp}$ and
$a_\mu^{SM}$: though a such agreement is quite good from $\tau$-decay 
(less than
1.1 $\sigma$), there  still remains some disagrement (1.5 to 3.3 $\sigma$)
between
$a_\mu^{SM}$ from $e^+e^-$ and $a_\mu^{exp}$.
More refined estimates of $a_\mu^{SM}$ require improved measurements of
the masses and widths of the scalar mesons, which are, at present, 
the major obstacles for
reaching a high-precision value of $a_\mu^{SM}$. These new measurements are
also necessary  for making progresses in our QCD understanding of the 
nature of scalar
mesons, which play a vital r\^ole on our understanding of the 
symmetry breaking in QCD.
In addition to some problems mentioned in
\cite{DAVIER} (absolute normalization of  the cross-section, 
radiative corrections,
$SU(2)$ breakings,...), the discrepancy between the $\tau$-decay and $e^+e^-$
data should stimulate improvements of the data and  experimental searches for
new and presumably tiny effects not accounted for until now. We plan to
come back to this point in a future work.
\section*{ACKNOWLEDGEMENTS}
\nin
Stimulating exchanges with William Marciano and Simon
Eidelman are gratefully acknowledged.

\end{document}